\documentclass[journal]{vgtc}                     % final (journal style)
\onlineid{0}

\vgtccategory{Research}

\vgtcpapertype{please specify}

\title{Towards Collective Storytelling: Investigating Audience Annotations in Data Visualizations}

\author{%
  \authororcid{Tobias Kauer}{0000-0003-0746-904X},
  \authororcid{Marian Dörk}{0000-0002-3469-7841}, and
  \authororcid{Benjamin Bach}{0000-0002-9201-7744}
}

\authorfooter{
  \item
  	Tobias Kauer is with the University of Edinburgh (UK) and the University of Applied Sciences Potsdam (DE).
  	E-mail: kontakt@tobiaskauer.org
  \item
  	Marian Dörk is with the University of Applied Sciences Potsdam (DE).

  \item Benjamin Bach is with INRIA (FR) and the University of Edinburgh (UK).
}

\abstract{%
This work investigates personal perspectives in visualization annotations as devices for collective data-driven storytelling.
Inspired by existing efforts in critical cartography, we show how people share personal memories in a visualization of COVID-19 data and how comments by other visualization readers influence the reading and understanding of visualizations.
Analyzing interaction logs, reader surveys, visualization annotations, and interviews, we find that reader annotations help other viewers relate to other people’s stories and reflect on their own experiences. Further, we found that annotations embedded directly into the visualization can serve as social traces guiding through a visualization and help readers contextualize their own stories. With that, they supersede the attention paid to data encodings and become the main focal point of the visualization.
  }


\teaser{
 \centering
 \includegraphics[width=\textwidth]{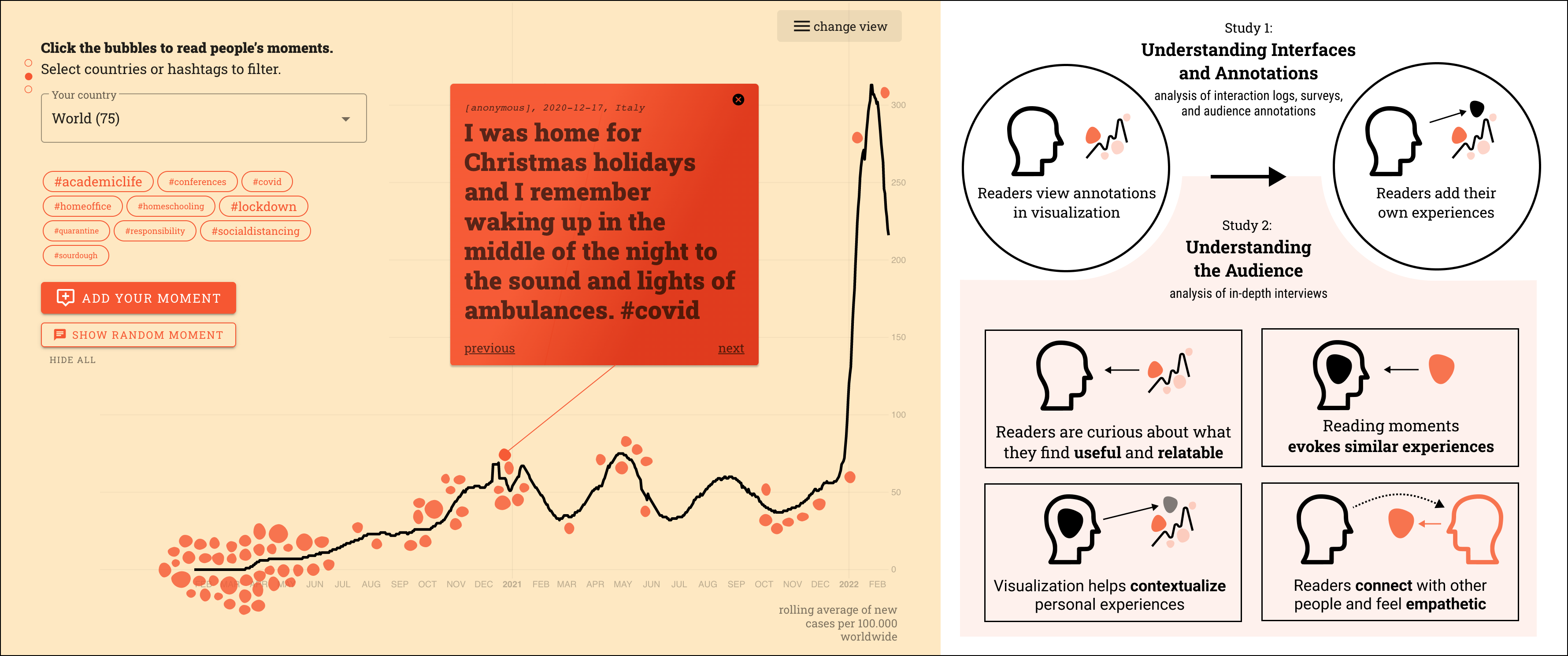}
 \caption{
\tool{} platform with embedded audience annotations (left) and their impacts on reading experience (right).}
 \label{fig:header}
}

\graphicspath{{figs/}{figures/}{pictures/}{images/}{./}} % where to search for the images

\usepackage{tabu}                      % only used for the table example
\usepackage{booktabs}                  % only used for the table example
\usepackage{lipsum}                    % used to generate placeholder text
\usepackage{mwe}                       % used to generate placeholder figures

\usepackage{mathptmx}                  % use matching math font

\usepackage{enumitem}
\usepackage[normalem]{ulem}

\DeclareRobustCommand{\statement}[2]{\textit{``#1''}~[#2]}
\DeclareRobustCommand{\statementWithoutParticipant}[1]{\textit{``#1''}}
\DeclareRobustCommand{\ben}[1]{{#1}}
\DeclareRobustCommand{\new}[1]{{#1}}
\DeclareRobustCommand{\newnew}[1]{{%\color{blue}
#1}}

\DeclareRobustCommand{\remove}[1]{{\color{red}}}
\DeclareRobustCommand{\add}[1]{#1}

\DeclareRobustCommand{\moment}[1]{{\sethlcolor{hideHighlight}\hl{\textit{``#1''}}}}
\DeclareRobustCommand{\moment}[1]{{\textit{``#1''}}}

\newcommand{\tool}{coronaMoments} %FOR ANONYMIZATION
\newcommand{\rannotations}{\textit{Q-Annotations}}
\newcommand{\rorigin}{\textit{Q-Origin}}
\newcommand{\rintegration}{\textit{Q-Integration}}

\newcommand{\vpersonalembedded}{\texttt{Audience\-Embedded}}
\newcommand{\vpersonaljux}{\texttt{Audience\-Separate}}
\newcommand{\vinstitutionalembedded}{\texttt{Author\-Embedded}}
\newcommand{\vinstitutionaljux}{\texttt{Author\-Separate}}
\newcommand{\viewa}{\vpersonalembedded}
\newcommand{\viewb}{\vinstitutionalembedded}
\newcommand{\viewc}{\vpersonaljux}
\newcommand{\viewd}{\vinstitutionaljux}

\DeclareRobustCommand{\revision}[1]{#1}
\DeclareRobustCommand{\revisiondelete}[1]{}
\begin{document}

%%%%%%%%%%%%%%%%%%%%%%%%%%%%%%%%%%%%%%%%%%%%%%%%%%%%%%%%%%%%%%%%
%%%%%%%%%%%%%%%%%%%%%% START OF THE PAPER %%%%%%%%%%%%%%%%%%%%%%
%%%%%%%%%%%%%%%%%%%%%%%%%%%%%%%%%%%%%%%%%%%%%%%%%%%%%%%%%%%%%%%%

\firstsection{Introduction}

\maketitle

Textual annotations have an important effect on people's understanding of data visualizations~\cite{kong2018frames}. 
Prior work has explored the crucial role that comments and text elements play for incorporating knowledge~\cite{lin2021data}, social analysis~\cite{heer2007voyagers,willett2011commentspace}, and narrative framing~\cite{hullman2011visualization}. 
However, while these works mainly focused on informative annotations provided by the visualization designer or story author, there is limited work~\cite{kauer2021public} about how annotations talking about \textit{personal stories} by the visualization readers
influence the reading experience of other readers. How would a plurality of 
reader annotations, personal perspectives, narratives, and insights influence the engagement with data and visualization?
\ben{Further, previous studies in human-computer interaction (HCI) and data visualization investigate how audiences share stories publicly and how this helps gather multiplicity of personal experiences, testimonies, or opinions~\cite{kauer2021public, dimond2013hollaback, maharawal2018anti}. 
One such format are interactive maps onto which personal anecdotes are shared as annotations that place them in a spatial context meaningful to other readers. 
Audience annotations on maps can raise awareness for different views, promote healthy behavior, and provide visibility of otherwise invisible stories \cite{kirby2021queering, maharawal2018anti}. 
The term \textit{collective storytelling} has been coined to refer to this way of sharing information
~\cite{dimond2013hollaback} and we see great potential for such shared stories and annotations for data visualization in general: providing a public platform to debate data representations, to vote on them~\cite{huron2021visualizations}, \newnew{to express personal evidence, formulate critique, and raise questions}~\cite{kauer2021public}.
In other words, visualization can become a public medium for collective engagement to promote humanized and critical perspectives on topics and their data within a data-driven discourse.}

\ben{In this paper, we contribute two studies about the influence of audience annotations on data visualizations and assess, to what extent existing efforts participatory storytelling can be translated to line charts.
The studies are informed by a design space for visualization annotations (Figure \ref{fig:designspace}). The first dimension describes the \textbf{origin of annotations}, i.e., whether annotations are provided by a visualization \textit{audience} or if they were provided by the visualization \textit{author}.  
The second dimension describes the \textbf{spatial integration} of annotations with a visualization, i.e., whether they are \textit{embedded} into a visualizations and reference specific visual marks or whether they are \textit{separated} from the visualization in a  comment thread. We conjecture that  embedded audience comments lead to more critical engagement with the visualization content.}

\ben{In our first study we designed an annotation interface---\textit{\tool{}}---\newnew{visualizing case numbers of the COVID-19 pandemic as a line chart. The interface allows to add annotations or browse author annotations, covering} three of the conditions from our design space.
Over the course of six months, we collected 77 public annotations, 949 interaction logs, and 62 survey responses. Our findings indeed show that embedded annotations are considered more engaging than separated ones and that audience annotations result in longer engagement times than author annotations.

In our second study, we asked eight visualization readers about their perceived 
benefits and impacts of each condition of our design space.
We found that embedded annotations change how data visualizations are framed and how other readers, in turn, engage with them.
\newnew{Audience annotations} are
\newnew{considered most useful to} readers,
\newnew{as they} allow them to relate to the visualization, evoke own experiences, and provide for a more meaningful reading experience. This has implications for the research and design of applications for data-driven storytelling, collective storytelling, and \newnew{surveying opinions from vast audiences}.}
\section{Background}\label{sec:background}
\add{Our research builds on prior work on critical approaches to data visualization and is inspired by existing efforts in participatory storytelling in cartography.}

\add{\subsection{Plurality in visualization}}
Donna Haraway disputes the idea that one can produce knowledge without being affected by one's personal history, partial perspectives, and cultural imprints~\cite{haraway2013situated}. Her critique of the ``view from nowhere'' has been applied to HCI, pointing out that data and data visualizations are situated in the particular contexts of those who record, create, and design them~\cite{d2020data}. 
Visualization design can deliberately try to hide the specific contexts, shortcomings, and limitations of its underlying data~\cite{hengesbach2022undoing}, akin to a ``one-way street'' shaped by the knowledge, ability, and desired narrative of the visualization author~\cite{d2015would}. 
Alternatively, researchers have called for a broader inclusion of voices in the design process~\cite{d2015would} as a condition for drawing a plurality of interpretations~\cite{dork2013critical}. 
While there are some examples that enable people to influence the narrative of a map~\cite{kirby2021queering, dimond2013hollaback, maharawal2018anti}, this effort has not been translated to other kinds of visualizations. 
Data visualization has been studied extensively in the context of social analysis~\cite{willett2011commentspace, heer2007voyagers}, but there are only few that concede power to the audience to shape the narrative of a visualization: \textit{DataHunches}~\cite{lin2021data} offers experts the ability to manipulate data representations based on their local knowledge. Similarly, \textit{viscussion}~\cite{kauer2024discursive} allows a wider audience to change the appearance of visualizations by visually anchoring critical, personal, and analytical discussions in it. However, prior work lacks a systematic review of the role of audience annotations: What is the impact of visual integration and does it matter whether annotations are created by the audience or by the visualization authors? This paper aims to  provide a better understanding of these factors, in order to determine the impact of audience annotations for plurality on visualization.

\new{\subsection{Crowdsourcing and mapping stories}} 
In cartography, a wide range of participatory data collection and mapping activities have spawned many projects\revisiondelete{ \newnew{in the tradition of citizen science and crowdsourcing}~\cite{orangotango2019}}. 
For example, the Anti-Eviction Mapping project has created an archive that collects stories of displaced residents in order to build community, raise awareness, and express solidarity with the victims~\cite{maharawal2018anti}. 
For this purpose, written stories of evictions are collected and placed on---digital and physical---maps that show the extent of the problem of housing injustice on a geographic scale while also sharing the stories of particular community members.
Similarly, Queering the Map\footnote{https://www.queeringthemap.com} invites users to share written stories that document LGBTQ2IA+ experiences~\cite{kirby2021queering}. Stories are submitted anonymously and appear as a marker on a map. Instead of clearly defining \newnew{an} audience, Kirby et al. coin the phrase ``stories for someone'', i.e., stories that do not address anyone specifically, but are written to add to an archive of experiences~\cite{kirby2021queering}. While there is no immediate political agenda, the mission statement is ``to generate affinities across difference and beyond borders---revealing the ways in which we are intimately connected''~\cite{kirby2021queering}.
\new{Dimond et al.~\cite{dimond2013hollaback} investigate Hollaback, a platform to crowdsource stories of victims of street harassment. The study shows how sharing their experiences on the platform helped the respondents regain some of the power they had lost during the incidents. The authors then introduce the notion of \textit{collective storytelling} as a practice of framing and confronting social challenges by those who experience them, not by social movement organizations or those in positions of power. 
The spatial anchoring of stories across these projects suits geographic representation},
which can also create spaces for people to meet, exchange ideas, and make claims about the world in co-located settings, as observed by Loukissas and Ntabathia in their studies of map rooms~\cite{loukissas2021open}. 
\add{In this paper, we study annotations in the context of another common reference space: time and temporal data.}

\new{\subsection{Annotations for context\remove{ and sensemaking}}}
\revision{Withholding relevant context from the visualization audience can have negative impacts on understanding and transparency~\cite{burns2022invisible}.}
Annotations in the form of textual comments can provide context for data represented by a visualization.
\revisiondelete{Annotations are widely used in data-driven storytelling to explain data, to draw the reader's attention to specific parts of the chart, and to provide information that can not be conveyed through visual means alone~\cite{riche2018data, ren2017chartaccent}. }\remove{They can include background information on the represented topic, the data sets used, how the visualization is to be read, or insights by the visualization author. Annotations may vary in length and scope: ranging from brief description to verbose treatise. They can refer to an entire visualization or only parts of it.} With growing awareness of the notion that data does \textit{not} speak for itself~\cite{d2020data}, there is greater sensibility towards the capacity of annotations to add not only context, but also a narrative framing to a visualization~\cite{hullman2011visualization,kong2018frames}.
\new{For example, Hullman et al.~\cite{hullman2015content} studied comments on news articles with data visualizations. The study found that non-expert audiences are less likely to discuss the presented data, but that they instead focus on how the data is framed \newnew{and criticize misleading presentations}.
While they enter the comment section without a clear intent for collaboration, 
\newnew{commenters} learn from others' observations and social interactions around news issues.}

\subsection{Personal engagements with data}
Peck et al.~\cite{peck2019data} show that people find those visualizations most useful to which they can personally relate based on their educational backgrounds, political affiliations, and personal experiences.
Kauer et al.~\cite{kauer2021public} further report on nine distinct types of reactions people express when commenting on visualizations on the social media platform Reddit such as observations, conclusions, hypotheses or testimonies. 
In \textit{testimonies}, people described their experiences with the represented data in their everyday life, disclose biographical information and personal relations to the topic in the data. The driving force behind testimonies was found to be people's willingness to add their own story as anecdotal evidence to heighten or alleviate the narrative framing of a visualization. /revision{With an increasing degree of personal  engagement, the audience is able to perform higher complexity tasks~\cite{mahyar2015towards}.}  In this paper, we leverage this \revisiondelete{motivation} to study these personal annotations from a visualization audience.

\begin{figure*}[htb]
 \centering
 \includegraphics[width=\textwidth]{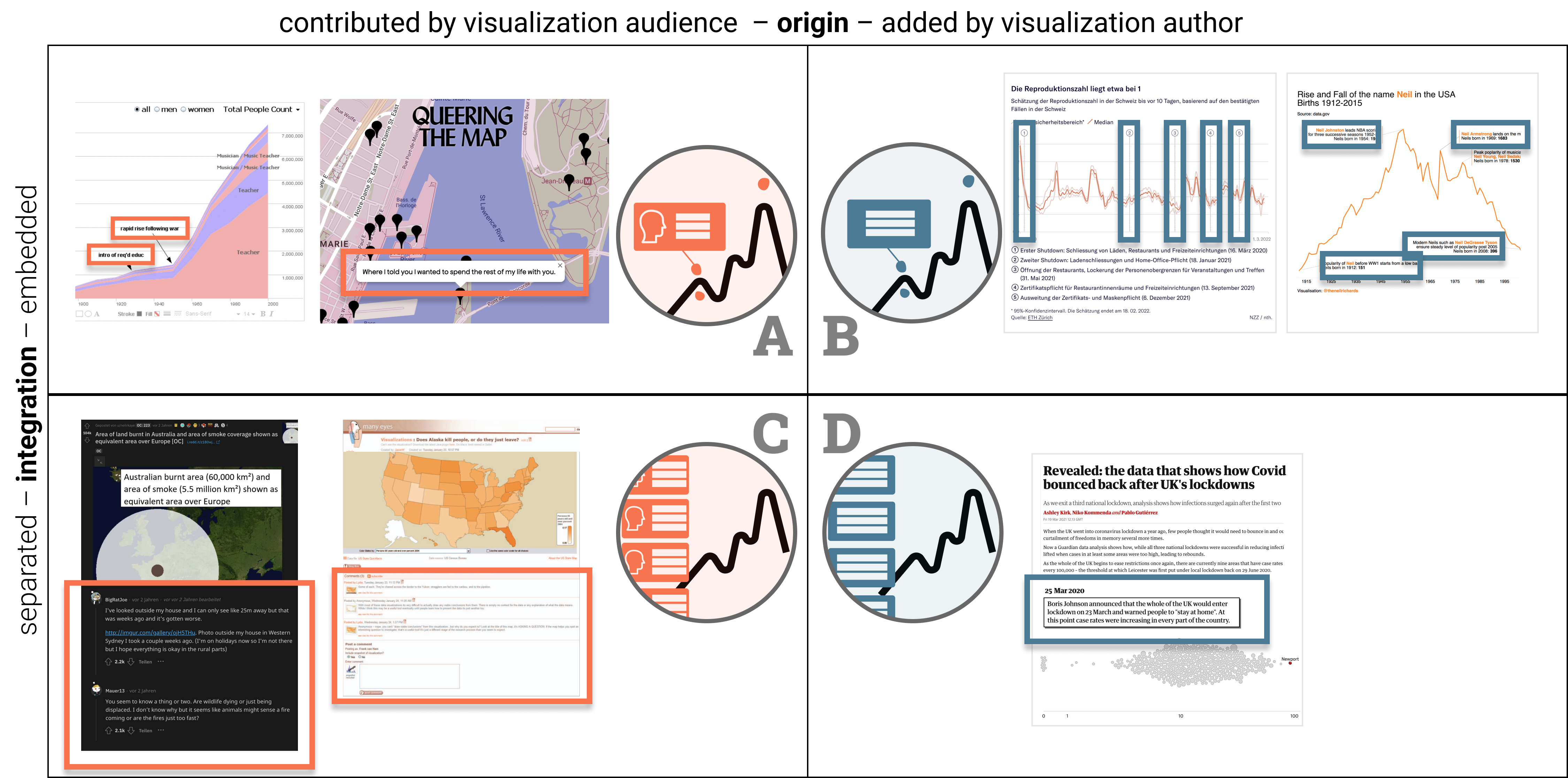}

 \caption{Our design space spans two dimensions: annotation origin and annotation integration. Each combination is illustrated by examples described in the Background, for each example the annotations are highlighted orange (audience annotations) or blue (author annotations): (A) annotations from the audience as investigated in social analysis~\cite{heer2007voyagers} or participatory mapping~\cite{kirby2021queering}; (B) embedded annotations curated by visualization authors as common in data-driven storytelling\textsuperscript{\ref{storytelling}}; (C) separate sections for comments from the audience in visualization-related articles or social media posts~\cite{hullman2015content, kauer2021public}, and (D) author's comments on the data that are separate from the visualization as often done in data journalism%\footnote{https://www.theguardian.com/world/ng-interactive/2021/mar/19/revealed-the-data-that-shows-how-covid-bounced-back-after-the-uks-lockdowns}.
}
 \label{fig:designspace}
\end{figure*}
\section{Approach}\label{sec:approach}

\add{Building on critical approaches to data visualization and participatory storytelling in cartography, we seek to examine how audience annotations within data visualizations can facilitate collective storytelling and influence viewer engagement. We focus on designing and evaluating a visualization interface that enables viewers to share personal annotations, situating their stories within the context of COVID-19 data.}
Our studies focus on the following three research questions.

\begin{itemize}[noitemsep,leftmargin=*,topsep=0pt]
    \item 
    \newnew{To get a understanding of people's anotations and to investigate the impact of the dimensions in our design space, we ask \textbf{[\rannotations]}:}
    \textit{
    How and why do audiences \revisiondelete{use personal annotations to }express themselves \revision{through annotations on}\revisiondelete{ in} public data visualizations?
    How and why do audiences express themselves through annotations on public data visualizations?
    }
    \item \newnew{Knowing that people express a broad range of comments when considering visualizations in public forums~\cite{kauer2021public}, we ask \textbf{[\rorigin]}: }\textit{How do annotations collected from the audience and annotations \revisiondelete{curated}\revision{added} by visualization authors differ in their effect on \revision{how they are read and understood}\revisiondelete{the viewing experience}?}
    \item \newnew{Participatory cartography shows stories in the place they are geographically situated in to provide a meaningful context~\cite{maharawal2018anti, kirby2021queering, dimond2013hollaback}. We lack understanding whether embedding annotations in a temporal context has a similar effect, and ask \textbf{[\rintegration]}: } \textit{How does embedding annotations into a visualization affect the viewer's experience?}
\end{itemize}

\subsection{Design space}\label{sec:designspace}
We propose a design space that is modelled after recurring themes in related work to better answer our research questions. On one dimension of that space, we illustrate the \textbf{origin} of an annotation, differentiating between annotations that are \revisiondelete{curated}\revision{added} by visualization authors and those that are sourced from the audience. On the other dimension, we illustrate the \textbf{integration} of annotations, which can either be embedded in the visualization or exist separately.
The resulting design space---illustrated in \autoref{fig:designspace}---informs the studies and discussion in the remainder of this paper.

\textit{\noindent\textbf{Dimension 1: Annotation origin}}---We refer to the origin of annotations through two main categories: \textit{Author annotations} refer to textual information \revisiondelete{provided and} \revisiondelete{curated}\revision{added} by the authors of the visualization\revision{. Such annotations can be provided by analysts, designers, or writers; be curated from existing (crowd-sourced) repositories; or be generated automatically by large language models.}
They are common in news media, infographics, data-driven storytelling and report on related events, people, facts, and insights from the data.
They are usually kept brief and neutral as they aim to provide an expert lens onto the data, helping to understand the visual encoding and visual patterns in a visualization as well as drivers, reasons, or consequences
around the data.
\textit{Audience annotations} on the other hand are reactions and reflections provided by the audience of a visualization by means of commenting. Audience annotations are dynamic in the sense that the audience can provide new annotations after a visualization is published. 
Audience annotations are common on news outlets, on social media \newnew{ platforms~\cite{hullman2015content,kauer2021public} or in social analysis~\cite{heer2007voyagers, willett2011commentspace}}

\textit{\noindent\textbf{Dimension 2: Annotation integration}}---The second dimension informing our research is the integration of annotations, by which we mean how annotations are visually and functionally situated with respect to the data visualization. \textit{Embedded annotations} are placed `inside' the visualization, eventually becoming part of the visualization. \revisiondelete{They are common in infographics or annotated charts~\cite{segel2010narrative} and could even include free-form pen-based annotations~\cite{satyanarayan2014authoring}.} \textit{Separated annotations} are placed juxtaposed to the visualization without any explicit reference to visual elements inside the visualization. Separate annotations are common in data-driven journalism, e.g., when insights from the visualization are discussed in the broader article, or in online communities with separate comment functions.

\begin{figure*}[htb]
 \centering
 \includegraphics[width=\textwidth]{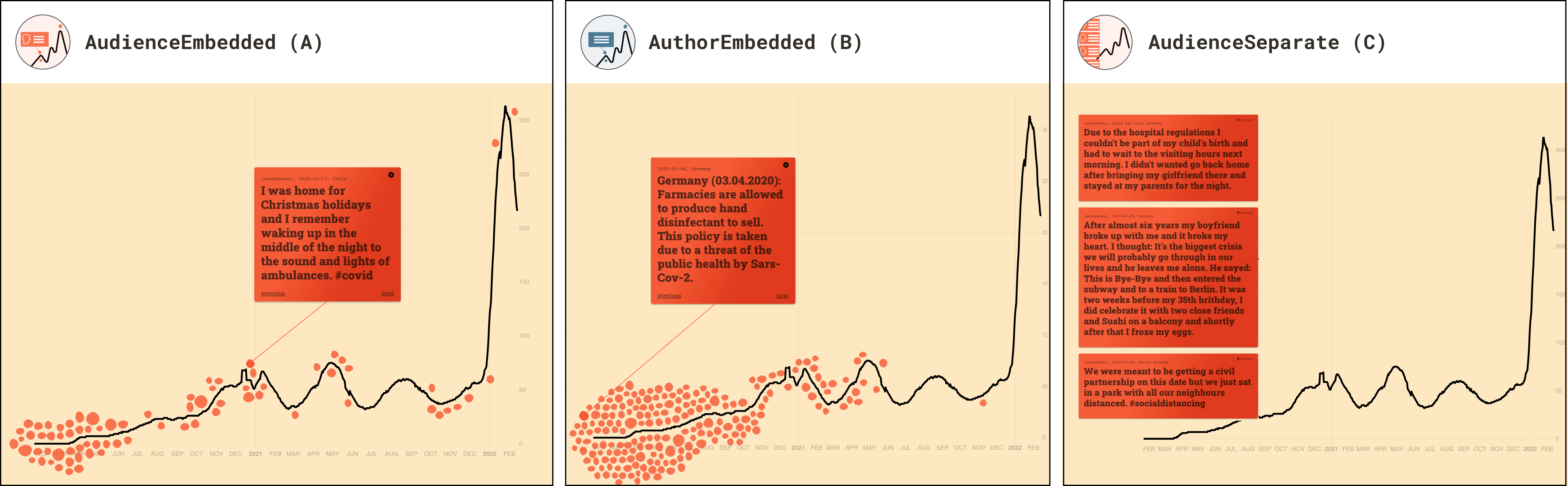}
 \caption{Three views of the \tool{} platform: (left) Embedded annotations collected from the audience, (middle) embedded annotations \revisiondelete{curated}\revision{added} by the author, (right) separated annotations collected from the audience.}
 \label{fig:three-views}
 \end{figure*}

\subsection{Study interface} 
\label{sec:tool}
\newnew{Our studies are situated within the context of the COVID-19 pandemic as the general public is familiar with the discourse, the data, and visualizations about the pandemic. The framing of visualizations in the context of the pandemic is a particularly sensitive topic~\cite{kong2018frames, lee2021viral}. There is a plethora of information available to curate annotations with meaningful connections to the data, e.g., timestamped information about government policies~\cite{cheng2020covid}.}
To run our studies, we built a tool called \tool{}. 
\new{The visualization shows a line chart depicting rolling averages of COVID-19 infections globally or for a selected country. We designed three viewing conditions~(\autoref{fig:three-views}) for our study, representative to three relevant quadrants in our design space (\autoref{fig:designspace}).}

\begin{itemize}[noitemsep,leftmargin=*,topsep=0pt]
\item \textbf{\viewa{} (\autoref{fig:three-views}a)}: This condition represents view A in \autoref{fig:designspace}; annotations are \textit{i)} provided by the visualization audience (origin) and are \textit{ii)} embedded directly into the visualization (integration). The view is representative for examples from participatory cartography~\cite{maharawal2018anti, dimond2013hollaback, kirby2021queering, loukissas2021open} that integrate audience annotations directly into the map canvas. Our version, shown in \autoref{fig:three-views}a, features orange dots---placed along the line chart---that represent the annotations left by the audience. Each dot is positioned based on its date and as close to the curve as possible.
Clicking a dot shows the full annotation in a pop-up within the visualization. The pop-up is connected to its respective dot with a line. Visitors can \revision{anonymously} place new dots in the chart to add their own annotations. The submission form suggests adding hashtags that were frequently used in other, already submitted annotations. \revision{Viewers could anonymously report potentially harmful comments for the authors to review, but no reports have been made.}
    
\item \textbf{\viewb{} (\autoref{fig:three-views}b)}: This condition represents view B in our design space; instead of audience annotations, it shows author annotations (origin) embedded into a visualization like in \viewa{} (integration). The condition represents examples in data journalism, in which visualization authors \revisiondelete{curate}\revision{add} annotations to provide context for temporal data.\footnote{{\label{storytelling}}https://www.storytellingwithdata.com/blog/2018/1/22/88-annotated-line-graphs}\textsuperscript{,}\footnote{\label{nzz}https://www.nzz.ch/panorama/coronavirus-in-der-schweiz-die-wichtigsten-grafiken-ld.1542774} Similar to \viewa~(A), this condition shows orange dots embedded into the charts. Instead of audience annotations, each dot represents a single policy response for the selected country, collected by the \textit{CoronaNet} research project \cite{cheng2020covid}. We use the project's existing categories as an equivalent to hashtags in personal stories. In this view, the audience can not add any annotations, as they are only \revisiondelete{curated}\revision{added} by the authors. 
    
\item \textbf{\viewc{} (\autoref{fig:three-views}c)}: This view refers to view C in our design space; audience annotations (origin) are shown as a scrollable list juxtaposed besides the visualization (integration).
The view is \revision{modeled after existing comment sections, e.g., on platforms like Reddit~\cite{kauer2021public}, news platforms~\cite{hullman2015content}, and prior work on annotations~\cite{willett2011commentspace}}
\revisiondelete{representative for existing interfaces in studies that investigate viewer reactions in separate comment sections}.
In our version (\autoref{fig:three-views}c), the list of annotations is scrollable and shows the same audience annotations as in \viewa{}~(A). The list is sorted chronologically. There are no orange dots embedded into the chart and no other visual relationship between the date mentioned in the comment and the chart. Visitors can add annotations through a button above and below the list which opens the same submission form as described in \viewa~(A). 
\end{itemize}

\noindent Besides these differences, all design variants have the same basic functionality: In \viewa{}~(A) and \viewc{}~(C), annotations can be filtered by hashtags provided by the audience (e.g., ``\#lockdown''). In \viewb~(B), annotations can be filtered based on existing categories from the data provider (e.g., ``Border Restrictions''). In both embedded views (\viewa~(A), \viewb~(B)), visitors can click a button to \textit{open a random moment} and can explore `\textit{next}' and `\textit{previous}' annotations based on their dates. In \viewc~(C), annotations are ordered by date.

\revision{We did not design a view condition for \viewd{}~(D) in our design space to increase the number of participants on views A-C, which are relevant to determine the role of embedded audience annotations.}
\section{Study 1: Placing annotations}
\label{sec:study1}

\revision{Before the first study, we piloted a first version of \tool{} with 25 members from two research labs. In this pilot, we only presented \viewa{}~(A), in order to collect an initial set of audience annotations to be displayed for the participants in the study to follow. We repeatedly invited participants in open calls via Slack and email to contribute moments to the platform and asked for feedback on usability. The pilot phase ran for two months, during which we collected 37 moments from a total of 25 lab members. To allow for a free expression of their experiences, comments were created anonymously. This also means that we have little insight into the comments author's demographics. As a result of the pilot phase, we decided to have additional surveys in the study to better undersand participants motivation}

\revision{After the lab-baed pilot, we desgined the first study} to be in-the-wild to obtain
answers from a broad and public audience. Since we were interested in people's intrinsic motivations to engage with the visualization and to contribute content, any
compensation or supervision would interfere with the validity of our findings.

\revisiondelete{fter piloting for usability, w}We publicly promoted the \textit{\tool} website through the authors' social media accounts and their research groups. 
We created a dedicated Twitter handle 
(@corona\_moments)
and infrequently posted screenshots of annotations to encourage site visits. \tool{} was online over six months, from October 2021 to March 2022.

\subsection{Study setup}
Upon visiting the website, each anonymous visitor was randomly assigned to one of our three viewing conditions and was asked to provide consent to participate in the study.
Prior studies have shown that \textit{intentional anonymity} enabled participants to respond more freely when expressing personal stories, desires, or fears~\cite{kirby2021queering}.
Upon consenting, participants in the \viewa{}~(A) and \viewc{}~(C) views saw a message asking \textit{``What stuck with us throughout this pandemic?''}. A short introductory sentence stated \textit{``Since the start of the pandemic we are confronted with charts about new cases daily. What are the stories behind the numbers?''}. In  \viewb{}~(B), we mentioned \textit{events} instead of \textit{stories}. We did not provide any further guidance as to what kinds of annotations people could submit to \textit{a)} avoid bias and \textit{b)} to encourage a wide range of annotations. 
After 20 seconds, viewers could manually switch to any of the other two view conditions through a menu at the top of the screen to explore those views, too. The delay was implemented to make sure that visitors would familiarize themselves with the assigned view first.

\subsection{Data collection}

We collected four kinds of information about the usage of the platform, the content of the shared annotations, and the motivation to participate in the study:
\begin{itemize}[noitemsep,leftmargin=*,topsep=0pt]
    \item \textbf{Interaction logs} 
    recorded timestamps of all interactions on the platform, including    
    viewing and submitting annotations.
    \item \textbf{Audience annotations} shared by visitors when viewing \viewa{}~(A) or \viewc{}~(B) were stored anonymously.
    \item A \textbf{general survey} 
    invited visitors to share what they found most interesting on the platform through a multiple-choice questionnaire (\autoref{fig:surveyresponses}). 
    \item A \textbf{post-submission survey} asked visitors in \viewa{}~(A) and \viewc{}~(C) who shared a moment about their motivation for the submission. 
\end{itemize}

\subsection{Findings}\label{sec:study1-findings}
Throughout the first study, 77 personal annotations were collected on the \tool{} platform: 37 during the pilot phase, and 40 more throughout the public deployment. To answer [\rannotations], we analyzed the collected annotations and subsequent survey submissions. The vast majority of annotations were submitted from Germany (69\%), followed by United Kingdom (12\%), three or less came from Italy, Norway, Netherlands, United States, Romania, Australia, Mexico, Greece, and Brazil. 

\subsubsection{\new{Collected annotations}}
\new{To provide a high-level understanding of the submitted annotations, we present selected comments that refer to recurring types of contributions. For example, we found many instances of people describing \textit{behavior changes} throughout the pandemic. Early on, people talk about how they did things for the last time (e.g., \moment{I attended my last in-person conference. News about the virus were spreading, but it all seemed far away back then. Everyone was making jokes about the new ways of greeting each other without shaking hands. \#conferences \#academiclife}). Later on, people describe their \textit{new normal}, which contemplates coping mechanisms to deal with the lockdown, including  baking bread, buying plants, going hiking, or buying home-office equipment. These contemplations also include  \textit{reflections} about how the pandemic affected their lives more broadly: \moment{Yesterday I realized that this pandemic has been going on for so long that I have become a different person in the meantime. There is no going back to my old self. I'm not sure about this new self either}. We further found many comments in which people describe situations they got into unwillingly, e.g., \moment{I went for a walk with a friend and we suddenly ended up in a small corona deniers / anti maskers demo without really noticing because they were walking in smaller groups. And then everywhere we wanted to go, paths were blocked by the police}. One prevalent element throughout the comments were notions of strong feelings, including anxiety (e.g., \moment{Catching my first common cold since the pandemic had started and I was *very* anxious about the symptoms until I had the negative result of the covid test}), joy (e.g., \moment{I adopted the sweetest black kitten and she is purrfect at keeping the loneliness bug away \#covidepet}), and anger and gloat (e.g., \moment{I called the police on a group of 40+ teenagers playing drinking games [...] It was pure joy seeing them busted}).}

\new{Generally, annotations describe moments that took place on a specific date. They rarely mention the number of COVID-19 cases that are represented in the chart on that date. Annotations have a meaningful relation to time
, but are only loosely connected to the data.
As instances of local knowledge~\cite{geertz2008local,lindblom1979usable}, they collectively reflect people's firsthand experiences of their changing social and physical environment during the pandemic.}

\begin{figure}[tb]
 \centering
 \includegraphics[width=0.5\textwidth]{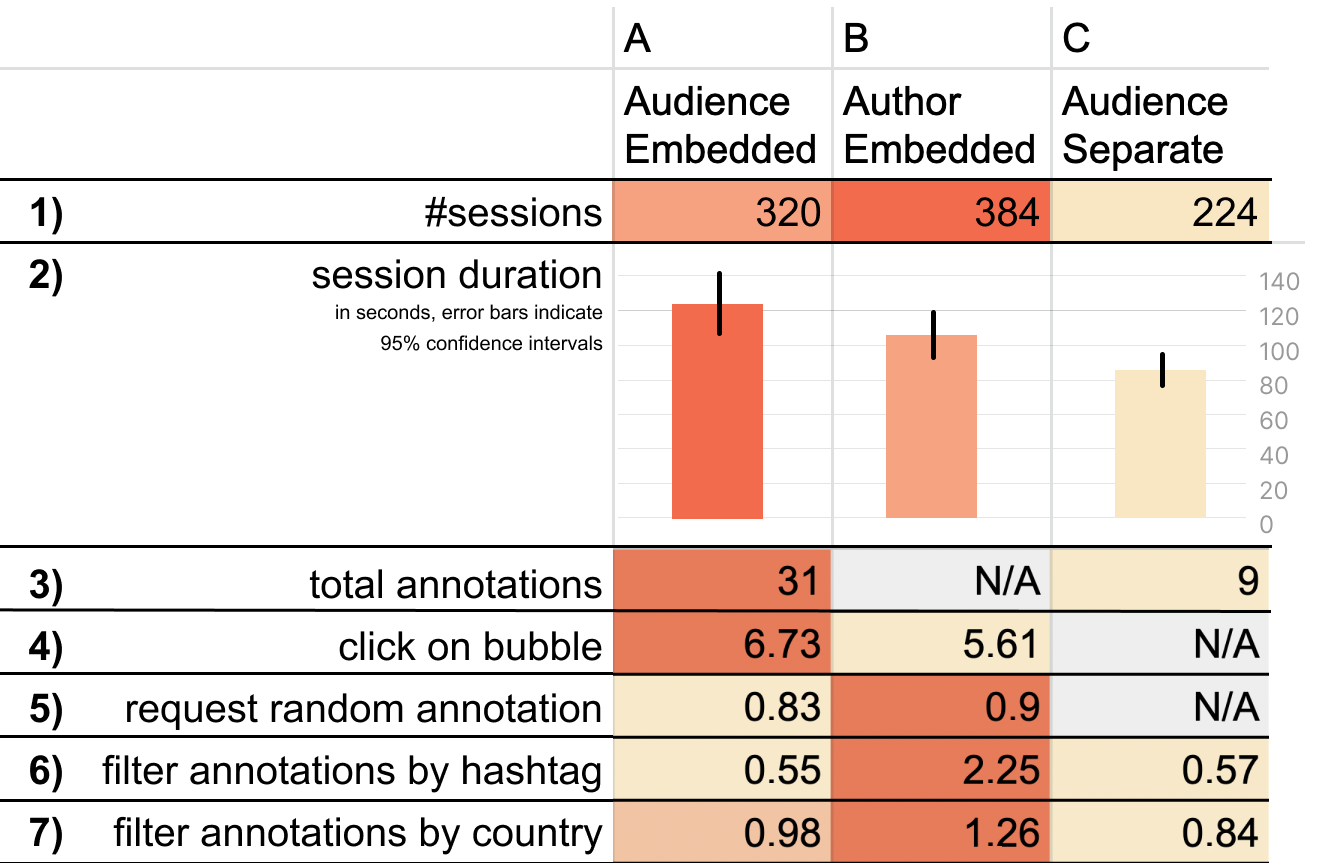}
 \caption{Across all conditions we compare, how many unique visitors viewed it (1), how long each session lasted (2), how many annotations they created (3), how often per session they enlarged annotation bubbles to read them (4), how many times per session they clicked the "show random moment" button (5), how often per session they filtered the view by selecting a hashtag (6) or a country (7).} 
 \label{fig:interactions}
\end{figure}

\subsubsection{Interactions}
%~\\
We collected data from 2409 unique sessions on the platform. We dismissed 259 sessions from the pilot phase \newnew{because it did only offer one view in order to collect an initial set of annotations}. Since we did not optimize all three views to be equally accessible on small devices, we dismissed 607 sessions of people who accessed \tool{} from a mobile phone. 596 people did not consent to the data collection; their sessions were hence removed. \revisiondelete{Based on t}\revision{T}he remaining 949 sessions\revisiondelete{, we could observe}\revision{indicate} several \revisiondelete{usage }patterns. \revision{To account for varying numbers of sessions across different views,  in figure~\ref{fig:interactions} we count interactions \textit{per session}.}
\begin{itemize}[noitemsep, leftmargin=*,topsep=0pt]

    \item \textbf{View visits}: \vinstitutionalembedded{} was visited most (40.2\%), followed by \vpersonalembedded{} (33.7\%), and \vpersonaljux{} (25.9\%). Based on the timespan between the first and last interaction per session, people spent the most time in \viewa{} (122 seconds avg.), followed by \viewb{} (104s avg.) and \viewc{} (84s avg.).
    
    \item \textbf{Read moments}: Similarly, visitors were more likely to click on and read audience annotations (avg. clicks per session: 6.7) than policy responses (5.6). Since in \viewc{} no clicks were necessary to read the moments, we cannot determine with confidence how many were read.
    
    \item \textbf{Filter moments}: The usage of UI elements to filter visible moments varies considerably across the views. In both audience views, few people filtered moments by hashtag (average 0.6 clicks per session), while this number quadruples for \viewb (2.3).  Filter by country shows a similar, yet less drastic distribution (\viewa: 1; \viewb: 1.3; \viewc: 0.8); so does opening random elements by clicking on the button ``Show Random Moment''  (\viewa: 0.8; \viewb: 0.9).
    
    \item \new{\textbf{Moment creation}: Visitors were more likely to submit a moment in \viewa{} (31, 78\%) than in the \viewc{}. Also, people are more likely to comment about the beginning of the pandemic (74\% of annotations refer to moments in 2020), irrespective of when people submit the annotation (58\% of annotations are posted at least one year after the moment they refer to).}
\end{itemize}

\subsubsection{Surveys}
\begin{figure*}[]
 \centering
 \includegraphics[width=1\textwidth]{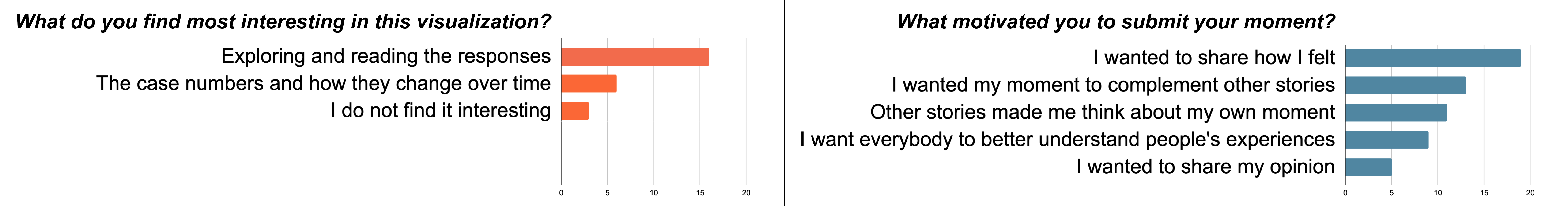}
 \caption{Survey responses of people who read annotations (red) and after they submitted own annotations (blue).}
 \label{fig:surveyresponses}
\end{figure*}

Out of 949 visitors, 24 (2.5\%) responded to the general survey shown on the website after a few minutes of usage. The responses indicate that people across all views are more interested in reading responses (66\%) than in the actual case numbers (25\%). 12\% of respondents expressed no particular interest in either (\autoref{fig:surveyresponses}). 38 visitors (4\%) responded to another survey \textit{after submitting an annotation} to give an account of their motivation. \revision{The form contained multiple choices (with multiple selections possible); the options participants could choose from are taken from Kauer et al.'s engagement drivers for user participation \cite{kauer2021public}: \textit{Expressing opinions}, \textit{matching knowledge}, \textit{sharing emotions}, \textit{claiming interest}, and \textit{adding anecdotal evidence}.} Responses indicate that the main driver for audience annotations was people's desire to share how they \textit{feel} (50\%) in contrast to sharing their \textit{opinion} (13\%). Other important motivations indicate the intricate relationship between people's own experiences and what they read about others: 
Reading other people's moments made respondents \textit{think about their own moments}~(29\%) which they then shared. They want their own stories to complement the existing collection of experiences~(32\%). Nine~(24\%) participants are well aware of the public display when submitting and wish for their moment to help others better understand their experiences. Similar to the first survey, only two participants~(5\%) focus on the actual graph. 
Both surveys included a free text input for participants to further explain their motivation and interest in viewing and sharing moments. While none of the participants of the first survey made use of that option, nine~(12\%) of the 77 participants who submitted a moment replied. Participants elaborated that their motivation for the submission was \statementWithoutParticipant{to share some current corona vibes}
and how it \statementWithoutParticipant{Feels good to reflect and write that down somewhere} and reflected on the impact reading other stories has on them: \statementWithoutParticipant{I found it touching to read other human experiences along time in the cases chart. It reminds me that I'm not alone despite the isolation and the social distancing}. 

~\\
Study 1 shows that embedded views are more engaging than separated views (\rintegration) and that audience annotations result in longer engagement times than author annotations (\rintegration).
\newnew{We found annotations that express people's experiences and emotions during the pandemic, generally without referencing specific data points represented in the chart.}
The log analysis shows that people spent the least amount of time in \viewc{}~(C) and were more likely to switch to any of the embedded views than vice versa. The effect of an annotation's origin is two-fold: On the one hand, showing audience annotations correlates with slightly longer sessions and more interactions. On the other hand, in \viewb{}~(B) people are more likely to engage with the visualization by filtering countries and categories. 
\section{Study 2: Experiencing annotations}
\label{sec:study2}
In our second study, we were interested in gathering further details about \textit{why} the three views were engaging in different ways and \textit{how} the audience annotations frame the engagement with the data. To that end, we conducted semi-structured interviews with eight participants around \tool{} with all the audience annotations from Study 1.

\subsection{Setup and participants}
We published an open call on the \tool{} website and shared it on Twitter, seeking potential interview candidates who are interested in the project to contact us via mail or direct messages. We recruited eight participants [P1-P8], all of whom had visited \tool{} at least once. Three participants identified as female, five as male; all participants had a bachelor's degree or higher. Participants came from a variety of fields, including the arts, HCI, journalism, and cultural studies.
Interviews were carried out online via Zoom and lasted 40--60 minutes. Participants were directed to the \tool{} website and accessed it on their computer. Each participant randomly started with one of the three views and could then switch to the other views, akin to study 1. Sharing their screen, we encouraged participants to freely explore each view for as long as they wanted. We watched participants using the platform and asked them to think aloud while addressing a set of four main tasks.

Due to the anonymous nature of our interaction logs, we did not know, whether each participant knew all views and the differences between them. At the beginning we asked an introductory question (\textit{``Can you explain to us, how the three views are different from each other?''}) that aimed at establishing a shared understanding of the visualization's functionality. We then encouraged participants to explore annotations in each view (\textit{``From each view, pick a response that stands out to you and explain why''}). Following Peck et al.'s approach to invite participants to share their values in data visualizations~\cite{peck2019data}, we asked participants to \textit{``Rank the three views according to how useful they are to you''}. In  follow-up questions, we asked  \textit{``What concept of usefulness do you have in mind when ranking the views''} and \textit{``How do different elements of the visualization (stories, policies and the chart on itself) help you reflect on the pandemic?''} During the interviews the first author took notes of the responses, revisited all afterwards, and grouped similar statements into general themes.

\newcommand{\arran}{\textsc{P1}}
\newcommand{\ryan}{\textsc{P2}}
\newcommand{\malte}{\textsc{P3}}
\newcommand{\marco}{\textsc{P4}}
\newcommand{\jo}{\textsc{P5}}
\newcommand{\christophe}{\textsc{P6}}
\newcommand{\lena}{\textsc{P7}}
\newcommand{\nicole}{\textsc{P8}}

\subsection{Findings}
We report key findings from the interview study regarding to usefulness of annotations, particularly their capacity to stimulate empathy and provide guidance for viewers to make sense of their own experiences.

\subsubsection{\textbf{Embedded annotations are most useful.}\label{embeddeduseful}}
\newnew{Most} participants ranked the three views of our study in the same order along descending usefulness: \viewa{}~(A), \viewb{}~(B), and \viewc~(C).
This ranking reflects our findings from study 1: people spend most time and read the most annotations when they are embedded in the visualization.
\newnew{Annotations in a separate view convey information, but do so without meaningful context}: \statement{It does give me insight, but I can't correlate it to the data}{\ryan}. This disruption between annotations and data is described as \statement{unproductive}{\marco} and impacted the perceived usefulness of \viewc{}~(C).

\subsubsection{\textbf{Personal relevance sparks curiosity.}\label{relevancecuriosity}}
To most participants, usefulness manifests itself in \statement{translating a data set in a way that it becomes relevant to the viewer}{\marco}. Relevance was most often described as personal interest with respect to personal experience and reflecting on one's own experience as well as learning about other people, comparing, and empathizing: \statement{What's important to me is really just my personal context and curiosity about other people}{\lena}. This observation is understandable, given the open nature of our study in the wild and the potentially many personal annotations from the audience related to the COVID-19 pandemic this research is situated in. Another reason for this focus on personal relevance might be the missing goal in engaging with the visualization and comments in the first place. Again, we think this is representative for many contexts in which visualizations are encountered, as opposed to a shared analysis goal in collaborative sensemaking. Reading personal annotations is not a means to an end, but satisfies a curiosity. For example, participants highlighted how clicking on embedded personal annotations sparked their desire for exploration and interest about other people's behaviors during the pandemic and how their stories relate to the data: \statement{What did people do in the first wave?}{\jo}. This includes comparing one self with other people and to understand how one's own choices, thoughts, and actions resemble those of other people: \statement{It's useful to compare my own experiences with others and to see how other people were doing compared to me}{\malte}. Participants especially appreciated their seemingly unique experiences to be reflected in the annotations shared by others: \statement{It's information that I can relate to. It's something that is unique to me}{\christophe}. Particularly, the personal character of the shared annotations offered a \statement{qualitative perspective that you can't really get from the data}{\ryan}.

Participants could also personally relate to annotations of government responses in \viewb~(B), e.g., regarding border closures: \statement{This one impacted me and it was a horrible one}{\arran}. Sometimes, the personal relation is not rooted in the policy's content, but in its writing style: \statement{this one speaks to me because it's so long and hard to understand, I have to read it twice. it reminds me of my situation, not knowing what is allowed and what is not}{\lena}.

\subsubsection{\textbf{Personal annotations evoke externalization. }\label{evokeexperiences}}
In all views, participants picked annotations they could personally relate to. For example, after reading personal stories, participants expressed how they had similar feelings: \statement{I read this and think to myself that I could have literally written it like that. It expresses exactly what I feel. It is good to know that I am not the only one who feels this way.}{\nicole}. Often participants have made comparable experiences, e.g., \statement{this person had to reschedule their wedding SEVEN times? I only had to do it twice}{\marco}, and throughout the interview they \statement{go down memory lane}{\marco} and share their own stories about cancelled conferences, closed libraries, their first COVID-19 tests, complicated travel plans and \statement{having similar discussions about similar topics}{\malte} when reading audience annotations. 
Reading other people's annotations inspired participants to create more annotations throughout the interview (\arran, \nicole). 

\subsubsection{\textbf{\revisiondelete{Visual}Distribution of annotations \revision{in visualization serving as} retrospective
\label{visualretrospective}}}
The participants pointed to the visual impression of all embedded annotations as a whole that makes a difference, compared to the separated view. They mention how the distribution of annotations in the visualization---many at the beginning of the pandemic and fewer towards the end
---represents their own experience of general fatigue: 
\statement{The frequency reflects my personal perception of the pandemic: in the beginning everything was new […]. Two years later, all of this is not so `interesting' anymore. One can understand why the entries are distributed like that}{\nicole}. Participants did not reflect on their own experiences just by looking at the visualization of incidence values and reading the lists of comments \textit{separated} from a visualization did not prompt similar reflections. It is the combination of data visualization and audience annotations integrated into a joint space
that seem to stimulate orientation, contextualization, and reflection on their own personal experiences and those of others. \new{Embedded annotations act as social traces that guide the viewers' attention to data and stories that helped them make sense of their own experiences.}

\subsubsection{\textbf{Contextualization of own experiences. }}\label{contextualizeexperience} 
When talking about the ways the interface helped participants with their own reflection, participants stated how the data representation supported them in looking back on the pandemic by setting mental anchor points: \statement{Having a chart support my memory helps recollecting memories at the right point in time}{\marco}.
The chart jumpstarted reflections about their memories and feelings \statement{the waves [in the chart] are meaningful in a way that they tell me how I felt in different months. They give me access to those points in time}{\jo}. Participants expressed how data representation helped them with their own contextualization: \statement{Looking back two years later, it is interesting to see what happened: How do I remember it, how do others remember it. For this, the timeline helps a lot}{\malte}. 
The visual distribution of annotations helps people reflect on the pandemic, the graph itself allows people to contextualize their own experiences within visually identifiable patterns of the data (e.g., surges in infections).

\subsubsection{\textbf{\new{Personal annotations can promote interpersonal connection}.}\label{empathy}} While browsing personal annotations, participants would empathize with other people's fate, even if they had not made similar experiences: \statement{That might not be something that relates to you - but might make you feel happy about them}{\ryan}. Personal annotations allowed the participants to change perspective and connect with other people through their shared experience: \statement{I have not heard this particular idea in my personal environment --- but now I feel connected to that unknown person!}{\nicole}. In two instances, people expressed how they take social distancing very seriously, but show leniency towards people who break rules, for example, to celebrate life events: \statement{I~can feel that. That person might go through so much deliberation before making such a decision}{\arran}. In these statements, participants express compassion for the authors of the audience annotations they encounter.
\section{Discussion}\label{sec:discussion}
According to our studies, viewers of public data visualizations are willing to share personal annotations, which in turn can lead to higher levels of engagements with the interface, mainly because they support reflection, orientation, and empathy.
\new{Furthermore, embedding annotations yielded longer engagements with the interface and a higher likelihood that a visitor would contribute an annotation.}
In the following, we answer our research questions, discuss findings, explore implications, and highlight important limitations. 

\subsection{Research Questions}
\textbf{\rannotations{}}:
In our studies, non-expert audiences use annotations not for collaborative analysis of data, but for personal expression. In contrast to existing work on commenting and collaborating in visualization~\cite{heer2007voyagers,willett2011commentspace}, the observed annotations may lack references to data, but provide a rich, subjective characterization of the time the data refers to. Prior classifications of audience reactions among other categories of expressions mention \textit{testimonies}: reports of first-hand experiences and anecdotal evidences that more often relate to a visualization's topic than do its data~\cite{kauer2021public}. \new{We find that annotations provide a thick description of people's local knowledge that contrasts the numeric data represented in the chart.}
The post-submission survey shows that the wish to express feelings is a strong motivator to share personal annotations. The process of creating annotations furthermore relies on other, existing annotations: reading annotations makes people think of their own stories, which they then can contribute as well.

\textbf{\rintegration{}:} 
Embedded views are seemingly more useful, create slightly longer sessions with increased engagement. While the \viewc{} view allowed participants to quickly skim all annotations, embedding annotations sparks a `desire to discover' and people are eager to explore annotations, particularly those from the audience. In contrast to existing interfaces that show annotations separately to the visualization (e.g., in online communities or on social media), embedded annotations help people navigate other people's stories along the temporal context provided by the data visualization and contextualize their own experiences, as well.
Furthermore, seeing the overall distribution of shared annotations within a data visualization can instill a sense of orientation in regard to the represented phenomenon echoing the ideas of social navigation during which collective traces guide users in digital information spaces~\cite{dieberger2000social}.

\textbf{\rorigin{}}: 
Audience annotations help making a data visualization personally relevant to the reader. They can promote empathy for other contributors and allow people to relate their own experiences to others---and with that, they play an important role for continuous engagement and participation. Audience annotations have great impact on the represented narratives and change what people think about after reading a visualization and what they are willing to contribute themselves. 
The personal character of the audience annotations seem to foster a qualitatively different connection between the reader, the visualization, and the represented phenomenon. 
However, the origin of annotations \textit{does not}
change the insights people draw from a visualization. 

\subsection{Limitations and reflections on methods}
Our work has some important limitations to consider.

\textbf{Participants:} We deliberately chose to only have anonymous submissions in order to allow for a greater freedom of expression. This means  we know little about the demographics of our contributors. While our experiment allowed a public audience to participate, there are still entry barriers: the website was only available in English and was not optimized for small screens, which excluded some demographics and usage scenarios. \new{Also, it is challenging to convince people to engage with content on the web in a voluntary study.} We primarily propagated the link among our own networks on Twitter, which limited our reach. Therefore, we might have missed annotations, survey responses and recordings of interaction logs from a more diverse audience who potentially engage differently with visualizations (e.g., on mobile phones), have had different experiences (e.g., members of high-risk groups), or find other things relevant. \revision{Further, the survey outcomes are based on only 24 responses for the first survey, and 38 responses for the second. This limits the external validity of the findings.}

The pilot phase of study 1 was run in two research labs and provided us with an initial set of annotations, which were necessary to have a meaningful comparison of different views. The limited diversity of the audience may have influenced the starting conditions of the experiment and follow-up annotations were potentially biased by existing ones. For example, the hashtag ``\#academiclife'' was introduced during the pilot phase and later automatically suggested to annotation creators during the public deployment. 
We do not know how responses would have differed with different starting conditions. \revision{Further, in different views (A and B) participants were confronted with different amounts of stories, which could have had an impact on engagement times.}

\textbf{Soliciting engagement:} The annotation submission was framed around \textit{stories} (``What are the stories behind the numbers?'') and invited people to ``share their \textit{moments}'', which potentially could have discouraged people from submitting a more diverse range of annotations. As outlined in Section 3, we focused on audience annotation as subjective and personal accounts, as opposed to more objective reactions provided by authors. Kauer et al.~\cite{kauer2021public} provide an overview of other possible audience reactions, such as \textit{clarifications}, \textit{opinions}, \textit{critique}, and more. Our work provides a better understanding of testimonies, but ignores other kinds of reactions. Future work should investigate different framings for soliciting engagement, for example, asking public audiences to embed annotations that express their critique of the data, commentary on the visualization design, or maybe even arguments on a social or political topic. 

\textbf{Data and topic:} We deliberately chose a topic people are well familiar with. Participants in our study however mentioned how they are tired of the pandemic, which may have affected general engagement and the amount of annotations created. We observed that annotations did not refer to specific data points, but talked about the topic more broadly. More research is needed to understand how other topics, data sets, and visualizations change audience annotations and their contextualization of a visualization.

\subsection{An open space for collective storytelling}
Based on our findings, we conceptualize visualization as a space that leverages audience annotations as mainstay for narrative framing and continuous participation. This space is \textit{open} to the general public to participate without pursuing explicit goals and allows for contributions of local, situated, and partial knowledge from the non-expert audience. It uses representations of data as a \textit{reference frame}, which helps the audience to contextualize their own knowledge and experiences in relation to the data as well as other people's annotations. In doing so, it enables a practice of \textit{collective storytelling}, in which the framing of a visualization does not consist of a singular narrative provided by the visualization authors, but instead is dispersed into a plurality of voices. Compared to data-driven storytelling, in which visualization authors control the narrative with their annotations, collective storytelling shifts this power to the audience, which can provide multiple narratives. We see five implications for visualization research and design.
    
\textbf{Rethink roles of annotations and visualization.} 
Data representations are typically the main protagonist of a visualization, while labels, captions, accompanying articles, and annotations are considered supporting structures that reinforce the message of the data. During the course of the presented research, we observed how these roles were gradually reversed and audience annotations became the primary concern of our visualization, while the chart was merely used for orientation\revisiondelete{ (See \ref{visualretrospective}, \ref{contextualizeexperience})}. With that, audience annotations have great potential to increase engagement and provide for a more profound reading experience. Prior work indicates that titles and captions have great impact on what insights one can draw from a visualization~\cite{kong2018frames}. Our findings suggest that audience annotations are similarly powerful devices that hold the potential to become the integral component of a visualization. Yet, more research is required to fully understand whether and how audience annotations can change what kinds of insights people draw from a visualization.

\textbf{Leverage personal relevance and curiosity.} 
\new{One challenge in visualization for social impact and civic issues is making the consequences of an issue relatable to the readers\revisiondelete{ in order to reduce the \textit{psychological distance}~\cite{slovic2010if}}}. Our study demonstrates how annotations from the audience established  personal relevance for other viewers and promoted empathy for the people writing annotations\revisiondelete{ (See~\ref{relevancecuriosity}, \ref{empathy})}. \revisiondelete{This has potential implications on future work for antropographics, asking whether providing human experiences as context in visualization has an impact on prosociality~\cite{morais2021can}.} We imagine embedded audience annotations as a future device in data visualizations for social movements that help viewers to relate to issues they learn about\revisiondelete{ (Section~\ref{sec:conclusion})}. 

\textbf{Cold-start contributions.}
The engagement with audience annotations requires a set of starter annotations that act as a seed for further contributions to be made. The data visualization alone may not be sufficient to foreshadow the kinds of experiences to be shared and stories to be told. It takes editorial effort to set the tone with the first set of statements. This is a daunting task. The \tool{} platform was originally called ``coronaMemories'', in hindsight, with an overly optimistic undertone that it might be over soon. However, especially when representing (and responding to) a highly contingent phenomenon, the interface---both visualization and annotations---turn into a dynamic, if not volatile, space that refers to a moving target. The uncertainty about the further development of the represented topic requires careful and open-ended wording both in the interface as well as in the seed contributions.
     
\textbf{Discourse among contributors.}
Our work considered annotations as singular responses that relate to a specific time. The findings from the studies suggest audience annotations are written in direct and indirect reference to other existing annotations\revisiondelete{ (See~\ref{evokeexperiences})}. Future interfaces can support the sharing of comments in response to annotations and with that support a multi-party discourse within a visualization. One challenge for such an interface is the visual arrangement of potentially multiple levels of annotations. We imagine these interfaces to reinforce the engaging effect of audience annotations, potentially helping to collect more annotations with a greater variety for further analysis.
    
\textbf{Prepare for moderation and manipulation.} 
With a potential for vast participation comes responsibility to prevent harm. Prior work showed how visualizations can be used to promote misinformation---specifically in a pandemic context~\cite{lee2021viral}. Audience annotations need to have robust mechanisms for reporting and moderation of potentially harmful content. The combination of quantitative evidence and qualitative experiences can prove particularly convincing, if not deceiving, when constructed in a manipulative manner.
There is a risk of misusing the authenticity and authority of crowd-sourced statements on a contentious issue. More research is needed to address questions of trust and transparency in data visualizations that are open to audience contributions. Depending on the topic and audience, having a plurality of potentially conflicting narratives in one visualization may confuse readers. Future work should investigate mechanisms for orientation while preserving the original multiplicity of voices from the audience, e.g., through options for self-categorization, filtering, or aggregation.
\section{A call to annotation}
\label{sec:conclusion}
We envision future visualization interfaces as open spaces for collective storytelling that allow audiences to engage with data, visualizations, and other readers \newnew{through} annotations. \newnew{Complementing prior work on democratizing visualization creation\revisiondelete{~\cite{viegas2007manyeyes}}, we see great potential for democratization of the discourse around visualizations.
To outline future opportunities in this space, we connect our findings to existing challenges in data visualization and other fields.}

Projects that aim to raise awareness for \newnew{data connected to} social issues, for example, can benefit from our findings that annotations promote personal relevance and interpersonal connection (See \ref{relevancecuriosity}, \ref{empathy}). Visualizations in the context of environmental and economic crises (e.g., climate change, biodiversity loss, inflation, housing crises) become more relatable when accompanied by a multiplicity of annotations that voice how people are affected. In these scenarios, raising awareness through crowdsourced personal annotations alone will not solve the problem, but it can provide coping mechanisms to people struggling to understand their own experiences~\cite{dimond2013hollaback}. \newnew{More research is needed to better understand whether stories that are not textual (e.g., \revision{drawings, }videos, photos, voice memos, drawings) have a similar effect and  and to assess whether increased personal relevance or interpersonal connection can contribute to a behavioral change.} 

Our finding that personal annotations evoke the submission of more personal annotations (See~\ref{evokeexperiences}) could also inform the design of future interfaces that support crowdsourced data collection for \newnew{data-driven policy making~\cite{veenstra2017data}. Local knowledge in the form of personal stories and people's frictions with their everyday environment is a powerful resource for stakeholders to make better decisions~\cite{geertz2008local}. Stories of people affected by data on economic hardship and related relief bills (e.g., student loans, stimulus packages, income support) can help evaluate and improve policy measures. \revisiondelete{Similarly, existing datasets of peace processes and their visualizations~\cite{bell2021ways} can benefit from integrating experiences of people on the ground. }More research and case studies are needed to understand how annotations can be well integrated into policymaking processes and to asses their trustworthiness and efficacy.}  

In the face of complex and dynamic topics that are publicly communicated in the form of data visualizations, we see great potential---but also risks---of visualization annotations for public discourse and deliberation.

\bibliographystyle{abbrv-doi-hyperref}

\bibliography{template}

\end{document}